\documentclass[12pt]{iopart}
\usepackage{iopams}
\usepackage{graphicx}% Include figure files
\DeclareGraphicsExtensions{.ps,.eps,.pdf}
\usepackage{bm}% bold math
\usepackage{bbm} %needed for \id to print as open face 1
\usepackage{xcolor}
\usepackage[normalem]{ulem}

\newcommand{\beq}{\begin{eqnarray}}
\newcommand{\eeq}{\end{eqnarray}}
\newcommand{\beqn}{\begin{eqnarray*}}
\newcommand{\eeqn}{\end{eqnarray*}}
\newcommand{\bra}[1]{\langle{#1}|}

\newcommand{\ket}[1]{|{#1}\rangle}

\newcommand{\ip}[1]{\langle{#1}\rangle}

\newcommand{\op}[2]{\ket{#1}\bra{#2}}

\newcommand{\id}{\mathbbm{1}}

\newcommand{\ii}{\ensuremath{\rmi}}
\newcommand{\dd}{\ensuremath{\rmd}}

\begin{document}

\title[Is coherence catalytic?]{Is coherence catalytic?}

\author{ Joan A. Vaccaro$^1$, Sarah Croke$^2$ and Stephen M. Barnett$^2$}

\address{$^1$ School of Natural Sciences, Centre for Quantum Dynamics, Griffith University, 170 Kessels Rd., Nathan 4111, Australia}
\address{$^2$ School of Physics and Astronomy, University of Glasgow, Kelvin Building, University Avenue, Glasgow G12 8QQ, United Kingdom}

\begin{abstract}
Quantum coherence, the ability to control the phases in superposition states is a resource, and it is of
crucial importance, therefore, to understand how it is consumed in use.  It has been suggested that
catalytic coherence is possible, that is repeated use of the coherence without degradation or reduction
in performance.  The claim has particular relevance for quantum thermodynamics because, were it true,
it would allow free energy that is locked in coherence to be extracted \textit{indefinitely}.  We address this
issue directly with a careful analysis of the proposal by \AA{}berg \cite{Aberg}.  We find that coherence
\textit{cannot} be used catalytically, or even repeatedly without limit.

\end{abstract}

\vspace{2pc}

\submitto{\JPA}

\section{Introduction}

Quantum coherence provides the ability to control the phases in superposition states and as such is
an essential element in the investigation and harnessing of quantum phenomena.  Indeed it is the
element that is at the very core of quantum phenomena, referred to by Feynman as ``the \textit{only}
mystery" \cite{Feynman}.  The thermodynamic significance of coherence has long been established, even
if not fully understood; indeed the link between masers (or lasers), which are the quintessential sources of
coherent light, and heat engines was made long ago \cite{Scovil,Scully}.

Coherence is a key component, perhaps the crucial distinguishing feature, of quantum thermodynamics
and it is essential, therefore, to have a reliable account of it as a resource.  Not to do so might lead to
at best inaccuracies and at worst, the prediction of phenomena that violate physical laws.

The idea of catalytic coherence and its variants has been applied to a range of topics including
the analysis of autonomous quantum machines \cite{Malabarba}.  In quantum thermodynamics,
\AA{}berg's repeatable property has been applied to the problem of extracting work from quantum
coherence \cite{Jennings} and, at a more formal level, the catalysis argument has been extended to
general symmetries \cite{Marvian}.  Recently, quantum catalysis has been employed in a study of
measurement-based quantum heat engines \cite{Hayashi}.  Quantum catalysis has become an
important part of the nascent field of quantum thermodynamics \cite{Goold}.
However, is widely acknowledged that for a system to act as a catalysis it must be returned to its initial state at the end of the process \cite{Du,Brando,Gallego,Ng2017}. It is readily apparent that this condition is not fulfilled in the processes treated by \AA{}berg \cite{Aberg}. Ng \emph{et al.} \cite{Ng2015} make a case for considering \emph{approximate} forms of catalysis on the grounds that all physical processes are approximate in some sense. But this misses the difference between what is possible in principle and technical limitations: even in the absence of technical limitations \AA{}berg's catalysis does not operate as a catalysis.

Here we present an analysis of a proposal by \AA{}berg that coherence is catalytic \cite{Aberg} or, perhaps
more accurately, that it is a resource that can be used repeatedly without degradation in performance
\cite{Jennings}.  We ask, specifically, whether the coherence in \AA{}berg's proposal is indeed
catalytic or repeatable and show that it is \textit{neither}.
In fact we show that coherence is a finite resource that is expended through use in accord with previous studies of the degradation \cite{BartlettNJP,BartlettJMO,BartlettRMP} and consumption \cite{White} of coherence.\footnote{ \AA{}berg used ``regenerating cycles'' to circumvent the loss of coherence attributable to the energy spectrum being bounded below. This kind of loss can also be circumvented by requiring the systems to be prepared in the upper energy state $\ket{\psi_1}$ and redefining the operator $U$ so that it gives $U\ket{\psi_1}=a\ket{\psi_0}+b\ket{\psi_1}$ in place of \eref{coherent operation}, where $\ket{\psi_0}$ is a lower energy state.
While this can reduce the overall loss in coherence, it does not eliminate the losses due to the inevitable correlations that build up between the source of coherence and the systems with each use, as we point out in detail below.}
Describing the use of coherence as catalytic, approximately catalytic, inexact catalysis or repeatable not only fails to capture this crucial property of coherence but suggests that the contrary is true.

We present a reanalysis of the \AA{}berg proposal concentrating, in particular, on the role of
correlations.  Our key finding is that the qubits to which the coherence is transferred are, necessarily
correlated and it is these correlations that limit the efficacy of repeated operations.  If we consider each
qubit independently then we do indeed find that they are in identical states but that these are
correlated.  In information theory it is common to speak of a sequence of systems being independent
and identically distributed (i. i. d.) \cite{Cover}.  For \AA{}berg's scheme the qubit states are indeed
identically distributed but they are not independent and so are not i. i. d..

To be completely clear, coherence is a strictly finite resource.  Repeated use inevitably degrades and ultimately
consumes it.  Once eliminated the residual coherence source performs no better than one prepared randomly.  In the
\AA{}berg proposal this is reflected in the \emph{complete destruction} of reservoir coherence following a \emph{single} and
ultimately inevitable error in the transfer of the phase reference to a qubit.

\section{Proposed scheme for catalytic coherence}

We begin with a brief presentation of the proposal by \AA{}berg for demonstrating catalytic coherence (CC) \cite{Aberg}.
The main idea explored in CC is exemplified by the use of a resource in the form of a multilevel quantum
system acting as an ``energy reservoir" that is initially in the coherent superposition state
\beq   \label{reservoir}
        \ket{\eta_{L,l_0}}=\frac{1}{\sqrt{L}}\sum_{l=0}^{L-1}  \ket{l_0+l} ,
\eeq
where $\ket{l_0+l}$ for $l\in \mathbb{Z}$ are reservoir energy eigenstates.
The coherence we seek to utilise is held in the relative phases between the amplitudes for the $L$ states
forming this superposition.  Here the phase is 0, but we could store a phase $\theta$ in the more
general state
\beq   \label{reservoir2}
        \ket{\eta_{L,l_0},\theta}=\frac{1}{\sqrt{L}}\sum_{l=0}^{L-1} e^{il\theta} \ket{l_0+l} .
\eeq
For simplicity we shall work with the state (\ref{reservoir}) but should keep in mind the fact that it is
being used as phase or coherence reference for $\theta = 0$.

We start with the general scheme but give, at the end of this section, a specific example, which might
make the scheme a little clearer.  The task we are required to perform is to prepare, repeatedly, coherent
superpositions of two-level systems (qubits), corresponding, at least approximately, to the operation
\beq  \label{coherent operation}
        U\ket{\psi_0}=a\ket{\psi_0}+b\ket{\psi_1}
\eeq
on a sequence of two-level systems, where $\ket{\psi_0}$, $\ket{\psi_1}$ are system energy eigenstates
and $U$ is a given unitary operator.  The coherent phase in the reservoir state, in particular, is imprinted
on the state as the relative phase of the amplitudes $a$ and $b$.

The process is analyzed in CC in terms of the quantum channels
\beq
        \label{channel Phi}
        \Phi_{\sigma,U}(\rho_0) =\tr_E[V(U)\rho_0\otimes\sigma V(U)^\dagger] \\
        \label{channel Lambda}
        \Lambda_{\rho_0,U}(\sigma) =\tr_S[V(U)\rho_0\otimes\sigma V(U)^\dagger]\ ,
\eeq
where $\tr$ denotes the trace operation.  Here $\Phi_{\sigma,U}(\rho_0)$ represents a channel that acts on system $S$ in state $\rho_0=\ket{\psi_0}\bra{\psi_0}$ and $\Lambda_{\rho_0,U}(\sigma)$ represents the complementary channel that acts on energy reservoir $E$ in state $\sigma$.  Here, the operator $V(U)$ acts on the tensor product of the associated Hilbert spaces ${\cal H}_S\otimes{\cal H}_E$ and is defined by
\beq    \label{defn of V(U)}
        V(U) = \sum_{n,n'=0,1}\ket{\psi_n}\bra{\psi_n}U\ket{\psi_{n'}}\bra{\psi_{n'}}
        \otimes\Delta^{n'-n}\ ,
\eeq
and $\Delta^k$, which is called the ``shift operator'', is defined by
\beq   \label{defn of Delta}
        \Delta^k=\sum_{j\in\mathbb{Z}}\ket{j+k}\bra{j}\ .
\eeq
Throughout we assume that $l_0$ in \eref{reservoir} is larger than the number of times the reservoir is used, so that
the interaction does not access the reservoir ground state, $\ket{0}$.  Hence we do not need to differentiate between
the doubly-infinite and half-infinite reservoirs, nor employ ``regenerating" cycles, as in CC\footnote{We note that the
regenerating cycles can also be avoided by (i) setting $l_0 = 0$ as for a half-infinite energy reservoir, (ii) requiring
the systems (qubits) to be prepared in the \textit{upper} energy state $\ket{\psi_1}$ before entering the channel, and
(iii) redefining the operator $U$ so that it gives $U\ket{\psi_1} = a\ket{\psi_0} + b\ket{\psi_1}$ in place of
\eref{coherent operation}.  Preparing the qubits in their upper energy state avoids the problem associated with
the reservoir having a ground state because interaction with each qubit can only increase the energy of the reservoir
or leave it unchanged when passing through the channel.}.

A key result of CC is that if $\tr(\Delta^a\sigma)\approx 1$ for $a=-2, \ldots, 2$ then
\beq   \label{approx U rho U}
   \Phi_{\sigma,U}(\rho_0)\approx U\rho_0 U^\dagger\ .
\eeq
Another key result is that the expectation value $\ip{\Delta^a}$ is invariant under the action of the channel on the reservoir $E$ in the sense that
\beq   \label{invariance of Delta}
        \ip{\Delta^a}=\tr(\Delta^a\sigma)=\tr [\Delta^a\Lambda_{\rho,U}(\sigma)]\ ,
\eeq
for all values of $a$.
These two results are the basis for arguing that the same channel can be used again on another system to perform the \emph{exactly} the same coherent operation, as epitomised explicitly in CC by the statement \cite{Aberg}
\beq   \label{same channel}
        \Phi_{\Lambda(\sigma),U}=\Phi_{\sigma,U}\ .
\eeq
This line of reasoning leads to the conclusion in CC that the coherence resource represented by the reservoir is not degraded by its use, and the claim that coherence has a \emph{catalytic} property as illustrated by phrases such as `coherence is catalytic in this model' and  `we only use the coherence catalytically and do not ``spend'' it at all' \cite{Aberg}.

Nevertheless, it is acknowledged in CC that the state of the reservoir $\sigma$ is changed by the channel $\Lambda_{\rho,U}$, i.e. $\Lambda_{\rho,U}(\sigma)\ne \sigma$.
This unavoidable change in the reservoir has prompted other authors to use alternative descriptors in place of \AA{}berg's `catalysis'.
For example, Korzekwa \emph{et al.} prefer to use `repeatable' to avoid any suggestion of an unchanging reservoir \cite{Jennings}.
Their argument is that for the channel to be repeatable, the reservoir only needs to remain as useful as it was initially irrespective of any change in its state.
A different choice is taken by Marvian and Lloyd who use the qualified description of `approximate catalysis' \cite{Marvian}.  To some extent the issue between these authors comes down to the meaning of the term
`catalysis'; this discussion, although of interest, is not the point of our paper.  For the interested reader, however,
we give a few historical remarks below\footnote{The word \textit{catalysis} is defined in the Pocket Oxford
Dictionary \cite{Oxford} as: \textit{Effect produced by a substance that, without undergoing change, aids chemical
change in other substances}.

The term ``catalysis" ({\it katalys} in the original Swedish) was introduced by
Berzelius \cite{Berzelius}.  A translation of his words given on the KTH website is \cite{KTH}:  {\it ``It is then shown that several simple and compound bodies, soluble and insoluble, have the property of exercising on other bodies an action very different from chemical affinity. The body effecting the changes does not take part in the reaction and remains unaltered through the reaction. This unknown body acts by means of an internal force, whose nature is unknown to us. This new force, up till now unknown, is common to organic and inorganic nature. I do not believe that this force is independent of the electrochemical affinities of matter; I believe on the contrary, that it is a new manifestation of the same, but, since we cannot see their connection and independence, it will be more convenient to designate the force by a new name. I will therefore call it the ``Catalytic Force'' and I will call ``Catalysis'' the decomposition of bodies by this force, in the same way that we call by ``Analysis" the decomposition of bodies by chemical affinity.''}

Catalytic processes have been known for a long time, although understanding their nature is a more
recent development.  It is interesting to note, however, that Sir Humphry Davy wrote on the topic and that this
was a significant element in the development of his famous safety lamp \cite{Davy}.}.

The presentation above is necessarily somewhat formal and an example calculation might be helpful.
Let us suppose that the desired transformation is $\ket{\psi_0} \rightarrow \ket{+} = 2^{-1/2}(\ket{\psi_0}
+ \ket{\psi_1})$.  The action unitary transformation acting on the first qubit and the energy reservoir
produces the state
\beq
V(U)\ket{\psi_0} \otimes \ket{\eta_{L,l_0}} = \frac{1}{\sqrt{2}}\left(\ket{\psi_0}\otimes\ket{\eta_{L,l_0}}
+ \ket{\psi_1}\otimes\Delta^{-1}\ket{\eta_{L,l_0}}\right) ,
\eeq
which is approximately the desired state.  To see this we can find the state of the qubit by tracing over the
energy reservoir to give the mixed state with density operator
\beq
\label{rhoSafter}
\rho_S = \left(1-\frac{1}{2L}\right)\ket{+}\bra{+} + \frac{1}{2L}\ket{-}\bra{-} ,
\eeq
where $\ket{-} = 2^{-1/2}(\ket{\psi_0} - \ket{\psi_1})$ is the state that is orthogonal to the desired superposition.
For large $L$ this is a very good approximation to the intended state.

The state of the energy reservoir following the interaction is changed
from the pure state $\ket{\eta_{L,l_0}}$ to the mixed state with density operator
\beq
\rho_E = \frac{1}{2}\left(\ket{\eta_{L,l_0}}\bra{\eta_{L,l_0}} + \Delta^{-1}\ket{\eta_{L,l_0}}\bra{\eta_{L,l_0}}\Delta\right) .
\eeq
This state has clearly changed, although the change is very small; the fidelity of the post-interaction state
with the initial state is
\beq
\bra{\eta_{L,l_0}}\rho_E \ket{\eta_{L,l_0}} = 1 - \frac{1}{L}\left(1 - \frac{1}{2L}\right) ,
\eeq
which is close to unity for large $L$.  The reservoir state has changed and in this sense the process is
not catalytic.  There are two senses in which the coherence appears to be catalytic and repeatable, however,
and this is the point: firstly, the post-interaction state of the energy reservoir is a mixture of two states,
$\ket{\eta_{L,l_0}}$ and $\Delta^{-1}\ket{\eta_{L,l_0}}$, each of which functions equally well as a source of coherence for
future use and secondly repeated uses of the energy reservoir as a coherence source to act on a sequence of
qubits will produce for each of them the \textit{same} mixed state (\ref{rhoSafter}).  This is the basis of the
claims for catalysis and repeatability, and it is these claims that we address in this paper.  We find, however,
that these promising indications are misleading.

\section{Independence versus quantum correlations}

We have seen that the \AA berg scheme creates qubits in the mixed state (\ref{rhoSafter}) but the single-qubit
state, which appears naturally in the channel picture, is only part of the story.  It is of the very essence of
``catalysis" or ``repeatability" that the coherence source should be used more than once, ideally many times.
A full description of the state of the qubits includes, not just the single-qubit properties, but also any correlations
that exist between them.  These correlations mean that the properties of a collection of qubits that have drawn
coherence from the reservoir are very different to those of uncorrelated qubits each in the state
$\rho_S$.  We demonstrate this point explicitly by considering first just two qubits, then a collection of $N$
qubits and contrast the properties of these with those of uncorrelated qubits.

\subsection{Two qubits}

We start by considering the action of our coherent transformation on a pair of qubits, each prepared
initially in the ground state $\ket{\psi_0}$.  Applying the unitary operation $V(U)$ to each in turn produces
the state
\beq
V(U)\otimes V(U)\ket{\psi_0}\ket{\psi_0}\ket{\eta} &=& \frac{1}{4}\left[\ket{+}\ket{+}(\id + \Delta^{-1})^2\ket{\eta}
\right. \nonumber \\
& & \quad  + (\ket{+}\ket{-} + \ket{-}\ket{+})(\id  - \Delta^{-2})\ket{\eta}   \nonumber \\
& & \qquad \left. + \ket{-}\ket{-}(\id  - \Delta^{-1})^2\ket{\eta}\right] ,
\eeq
where $\id $ is the identity operator and we have, for brevity, written $\ket{\eta}$ for the reservoir state
and omitted the tensor-product symbols where there is no ambiguity.  Here $V(U)\otimes V(U)$ is a
short hand for
\beq
V(U)\otimes V(U) &=& \sum_{n,n',m,m'=0,1}\ket{\psi_n}\bra{\psi_n}U\ket{\psi_{n'}}\bra{\psi_{n'}}\otimes  \nonumber \\
& & \qquad \qquad \ket{\psi_m}\bra{\psi_m}U\ket{\psi_{m'}}\bra{\psi_{m'}}\otimes\Delta^{n'-n+m'-m} .
\eeq
The resulting state of the two qubits is not separable and, in particular, is not
simply $\rho_S\otimes\rho_S$.  As a simple demonstration of this we give the probabilities for the outcomes
of measurements on the two qubits in the $\{\ket{+}, \ket{-}\}$ basis.  We find these to be
\beq
P(+,+) &=& 1 - \frac{3}{4L}  \nonumber \\
P(+,-) &=& \frac{1}{4L} = P(-,+) \nonumber \\
P(-,-) &=& \frac{1}{4L} ,
\eeq
where we have used the expression
\beq
\bra{\eta}\Delta^a\ket{\eta} = 1 - \frac{|a|}{L}  \qquad |a| \leq L .
\eeq
That there are correlations between the two qubits is clear from the fact that these probabilities do not
factor into products.  For comparison we give the products of the single-qubit probabilities:
\beq
P(+)\times P(+) &=&  1 - \frac{1}{L} + \frac{1}{4L^2}  \nonumber \\
P(+)\times P(-) &=& \frac{1}{2L} - \frac{1}{4L^2}  = P(-)\times P(+) \nonumber \\
P(-) \times P(-) &=& \frac{1}{4L^2} .
\eeq
These are the probabilities that would result if the channel picture were sufficient to describe two uses of the phase
resource so that the two-qubit state was $\rho_S\otimes\rho_S$.

The number of reservoir energy eigenstates involved is intended to be large, so we can take the large $L$
limit of these probabilities.  It is clear, in this limit, that on most occasions measurements of the two qubits
will result in the value `$+$', but it is when one or more `$-$' value occurs that we see the significance of the
correlations.  In the absence of correlations, the probability for getting two `$-$' outcomes is very small,
$\sim L^{-2}$, but the \AA berg scheme produces this outcome with a much higher probability, $\sim L^{-1}$.
Indeed it is noteworthy that all three outcomes in which at least one `$-$' occurs have the same probability.
This reflects a general feature on the correlations in the \AA berg scheme.  To see this clearly we consider
the properties of a larger number of qubits.

\subsection{$N < L$ qubits}

The correlations evident on our analysis of two qubits are yet more apparent and significant when we consider
a larger number of qubits.  For $N$ qubits (where $N<L$) the interaction produces the combined qubit-reservoir
state
\beq
V(U)^{\otimes N}\ket{\psi_0}^{\otimes N}\ket{\eta} =
\left[\frac{1}{2}\left[ \ket{+}(\id  + \Delta^{-1}) + \ket{-}(\id  - \Delta^{-1}) \right]\right]^{\otimes N}\ket{\eta} .
\eeq
From this general expression we can extract the probabilities that a measurement on each of the $N$ qubits
in the $\{\ket{+}, \ket{-}\}$ basis will give any chosen sequence of `$+$' and `$-$' results.  The symmetry of
the process means that the probability for any given sequence in which $n$ qubits are found in the state
$\ket{+}$ and $N-n$ in the state $\ket{-}$ is
\beq
P_{\rm seq}(n) = \frac{1}{2^{2N}}\bra{\eta}(\id -\Delta)^{N-n}(\id +\Delta)^{n}
(\id +\Delta^{-1})^{n}(\id -\Delta^{-1})^{N-n}\ket{\eta} .
\eeq
We emphasise that this probability does not depend on the order in which the qubits appear in this sequence
as $\Delta$ commutes with $\Delta^{-1}$.  This means, in turn, that the probability of finding
$n$ of the qubits in the state $|+\rangle$ in \emph{any order} is
\beq
P(n) = \left(\begin{array}{c}
					  N\\
					  n
					  \end{array}\right)P_{\rm seq}(n) \, ,
\eeq
and hence that the probabilities sum to unity, as they should:
\beq
\sum_{n=0}^N P(n) &=& \sum_{n=0}^N
            \left(\begin{array}{c}
					  N\\
					  n
					  \end{array}\right)
                 2^{-2N}\langle \eta|(2\id  - \Delta - \Delta^{-1})^{N-n}(2\id  + \Delta + \Delta^{-1})^n|\eta\rangle
\nonumber \\
&=& \langle \eta|\left(\frac{\id }{2} - \frac{\Delta}{4} - \frac{\Delta^{-1}}{4}
+ \frac{\id }{2} + \frac{\Delta}{4} + \frac{\Delta^{-1}}{4}\right)^N|\eta\rangle  \nonumber \\
&=& 1 \, .
\eeq
Finding a single qubit in the state $|-\rangle$ leaves the reservoir in a state that is essentially devoid of
the initial coherence and this suggests that the next qubit tested is \emph{equally likely} to be found in
the state $|-\rangle$ as in the state $|+\rangle$.  This means to suggest, in particular, that
\begin{equation}
P_{\rm seq}(N-1) = P(0) \, ,
\end{equation}
so that, for example, if the first (or any other) qubit is found in the state $|-\rangle$ then the remaining
$N-1$ are equally likely to all be found in the $|-\rangle$ or the state $|+\rangle$!  The reason for this
remarkable result is readily understood in terms of the state of the reservoir following a $\ket{-}$ outcome.
In this case the reservoir state, $\ket{\eta}$, is acted on by $\id - \Delta^{-1}$ and hence the (unnormalised)
reservoir state becomes
\begin{equation}
(\id-\Delta^{-1})\ket{\eta} = \frac{1}{\sqrt{L}}\left(\ket{l_0 + L - 1} - \ket{l_0 - 1}\right) .
\end{equation}
The are no adjacent or even nearby energy states in this case and hence it no longer acts as a source of coherence.  Preparation from it of a $\ket{+}$
or a $\ket{-}$ state will happen with equal probability.  More generally, the probability that the remaining
qubits form a given sequence with $m$ qubits in the state $\ket{+}$ and $N-m-1$ in the state $\ket{-}$ is the
same as that for a sequence in which $m$ qubits are found in the state $\ket{-}$ and $N-m-1$ in the state $\ket{+}$.

Evaluating the general probabilities $P_{\rm seq}(n)$ is a lengthy and not especially enlightening procedure,
but we have found excellent approximations to these, which give probabilities to within a few percent or better
for $N > 1$.  A few examples will suffice to indicate the trend:
\beq
P_{\rm seq}(N) &=& P(+, +, \cdots , +)  \nonumber \\
&\approx& 1 - \sqrt{\frac{N}{\pi}}\frac{1}{L}   \nonumber \\
P_{\rm seq}(N-1) &=& P_{\rm seq}(0) = P(0) \nonumber \\
&\approx& \frac{1}{2\sqrt{\pi(N-1)}L}  \nonumber \\
P_{\rm seq}(N-2) &=& P_{\rm seq}(1)  \nonumber \\
&\approx&  \frac{1}{4\sqrt{\pi(N-2)^3}L}  \nonumber \\
P_{\rm seq}(N-3) &=& P_{\rm seq}(2) \nonumber \\
&\approx& \frac{3}{8\sqrt{\pi(N-3)^5}L}  .
\eeq
The most striking feature of these probabilities is that those for which there is at least one `$-$' outcome
\emph{all} fall off as $L^{-1}$.  This contrasts strongly with the situation that \emph{would} hold in the absence
of the correlations, with the state $\rho_S^{\otimes N}$, for which $P_{\rm seq}(N-k)$ would fall off as $L^{-k}$.
The overall probability that there will be $N-n \ll N$ `$-$' outcomes is rather flat:
\beq
\label{Eq25}
P(N-1) &=& NP_{\rm seq}(N-1) \approx \sqrt{\frac{N}{\pi}}\frac{1}{2L} \nonumber \\
P(N-2) &=& \frac{N(N-1)}{2}P_{\rm seq}(N-2) \approx \sqrt{\frac{N}{\pi}}\frac{1}{8L}  \nonumber \\
P(N-3) &=& \frac{N(N-1)(N-2)}{6}P_{\rm seq}(N-3) \approx \sqrt{\frac{N}{\pi}}\frac{1}{16L} ,
\eeq
where we have simplified these expressions by choosing $N \gg 1$.  In the absence of these correlations,
with the multi-qubit state $\rho_S^{\otimes N}$, the situation is very different and for $N \ll L$, it is most
unlikely that more than one of the qubits will be found to be in the state `$-$':
\beq
P_{\rho_S^{\otimes N}}(N-1) \approx \frac{N}{2L}  \nonumber \\
P_{\rho_S^{\otimes N}}(N-2) \approx \frac{1}{2}\left(\frac{N}{2L}\right)^2  \nonumber \\
P_{\rho_S^{\otimes N}}(N-3) \approx \frac{1}{6}\left(\frac{N}{2L}\right)^3 .
\eeq
For $N \ll L$ only the first of these is comparable to the probabilities for a small number of `$-$' outcomes
given in Eq. (\ref{Eq25}).  The most extreme case is the probability that all the $N$ qubits will be found
in the state $\ket{-}$ which, as we have seen, is approximately $(2L\sqrt{\pi N})^{-1}$, while for the uncorrelated
state this probability has the vastly smaller value of $(2L)^{-N}$!

The correlations between the transformed qubits are a crucial part of the overall picture and although each
qubit, when considered alone, will be found in the state $\rho_S$, the multi-qubit state is \textit{very different} from the
uncorrelated tensor product of these density operators.  Multiple coherent operations, acting on multiple qubits
is the very essence of catalysis and repeatability, and it follows that these correlations cannot be ignored.
Neglecting these correlations can lead to unphysical conclusions as we demonstrate in the next section.

\section{Paradoxical repercussions}

The purpose of this section is to highlight the fundamental necessity for the existence of the correlations
we have described and, in doing so, expose the inadequacy of describing each post-interaction qubit
by the simple mixed state $\rho_S$.  This is important as it shows that the requirement that we account
fully for the correlations between the qubits is general and not simply a particular manifestation of the
\AA berg scheme.

\subsection{Unphysical state discrimination}

Our first example is one of quantum state discrimination \cite{Chefles,SarahRev}.  The key idea is
that it is not possible,
even in principle, to determine for certain in which of two known non-orthogonal quantum states a system
has been prepared.  The absolute minimum probability of error in making this choice is given by the Helstrom
bound \cite{Helstrom,Stevebook}.

Consider an energy reservoir to have been prepared in one of two possible initial states, $\ket{\eta(\theta_1)}$
or $\ket{\eta(\theta_2)}$ where
\beq   \label{reservoir theta}
  \ket{\eta(\theta)}=\frac{1}{\sqrt{L}}\sum_{l=0}^{L-1} e^{\ii l\theta}\ket{l_0+l} .
\eeq
In general these two possible reservoir states will not be orthogonal\footnote{The exception being only if
$\theta_1 - \theta_2$ is an integer multiple of $2\pi/L$} and if they are not orthogonal then it necessarily
follows that we cannot discriminate between these two states with certainly.

Let us suppose that the energy reservoir is used to prepare a very large number of qubits, each of which
will then be found in one of the mixed states
\beq
\label{Eq28}
 \hspace{-5mm}\rho(\theta_j) = \frac{1}{2}\left[\ket{\psi_0}\bra{\psi_0}+\ket{\psi_1}\bra{\psi_1}
        +\left(1-\frac{1}{L}\right)(e^{-\ii\theta_j}\ket{\psi_0}\bra{\psi_1}
        +e^{\ii\theta_j}\ket{\psi_1}\bra{\psi_0})\right] ,
\eeq
where $j = 1,2$.  If we accept literally the claim of CC that the same reservoir can be used repeatedly  to perform the same coherent operation and so create the state $\rho(\theta_j)^{\otimes N}$ then we can recast the problem of
determining the reservoir state as one of discriminating between the two $N$-qubit states,
$\rho(\theta_1)^{\otimes N}$ and $\rho(\theta_2)^{\otimes N}$.  The probability of error in discriminating between
these two states decreases with each additional copy available, and approaches {\it zero} in the limit of large $N$.
To show this explicitly, we note that the minimum achievable probability of error in discriminating two states $\rho$ and $\sigma$ is given by the well-known Helstrom bound \cite{Helstrom,Stevebook}:
\beq
        P_{\rm err} (\rho, \sigma) &= \frac{1}{2} \left( 1 - D(\rho,\sigma) \right)
\eeq
where $D(\rho,\sigma) = \frac{1}{2} \tr|\rho - \sigma|$ is the trace distance. Further, a bound on the trace distance is given by $D(\rho,\sigma) \geq 1 - F(\rho,\sigma),$ where
$F(\rho,\sigma) = \tr \sqrt{\rho^{1/2} \sigma \rho^{1/2}}$ is the fidelity \cite{Nielsen&Chuang}, thus
\beq
        P_{\rm err} \leq \frac{1}{2} F(\rho,\sigma)\ .
\eeq
For the $N$-copy states corresponding to different reservoir states the fidelity is readily calculated to be:
\beq
        F(\rho(\theta_1)^{\otimes N},\rho(\theta_2)^{\otimes N}) &= F(\rho(\theta_1),\rho(\theta_2)])^{N}\nonumber\\
        &=\left[1-\frac{1}{2} \left(1-\frac{1}{L} \right)^2 \left(1-\cos(\theta_1-\theta_2) \right) \right]^{\frac{N}{2}},
\eeq
which tends to zero exponentially as $N$ increases. It would appear, therefore, that the channel could be used to discriminate between two non-orthogonal reservoir states $\ket{\eta_{L,l_0}(\theta_1)}$, $\ket{\eta_{L,l_0}(\theta_2)}$ with an accuracy approaching 100\% \cite{Chefles,state separation}.  But this {\it contradicts} the fundamental result that no quantum measurement can unambiguously distinguish between two non-orthogonal states \cite{Stevebook,Nielsen&Chuang}.  Hence, we are left with a paradox: the results of CC---and \eref{approx U rho U}, \eref{invariance of Delta} and \eref{same channel} in particular---appear to imply that the channel $\Phi_{\sigma,U}$ can perform coherent operations repeatedly, and yet we have just seen that this possibility would lead to a violation of a fundamental result in quantum measurement theory.  The resolution, of course, lies in the correlations between
the qubits that are neglected in the channel picture.

\subsection{Unphysical generation of unbounded coherence}

Our second example raises the issue of quantum coherence as a limited resource and so challenges directly
the idea of its catalytic use.  We start by noting that
the coherence represented by the reservoir state in \eref{reservoir} is an example of a broken U(1) symmetry, and its coherence is quantified by its \emph{asymmetry} with respect to the U(1) symmetry group.
The asymmetry quantified by $A_G(\varrho)$ was first introduced by one of us \cite{VAWJarxiv,VAWJ} as a measure of the ability of a system with density operator $\varrho $ to act as a reference and break the superselection rule (SSR) associated with a symmetry described by the group $G$.
It is defined as \cite{VAWJarxiv,VAWJ}
\beq
        A_G(\varrho) \equiv S({\cal G}_G[\varrho ])-S(\varrho )
\eeq
where $S(\varrho )\equiv-\tr( \varrho  \ln \varrho )$ is the von Neumann entropy of the density operator $\varrho $ and ${\cal G}_G[\cdot ]$ is the twirl superoperator is given by
\beq   \label{twirl}
        \mathcal{G}_G[\varrho ]=  \frac{1}{|G|}\sum_{g\in G} T_g\varrho  T_g^\dagger
\eeq
for the unitary representation $\{T_g:g\in G\}$ of a discrete group $G$ of order $|G|$.
For continuous groups, the sum in \eref{twirl} is replaced with an integral with an appropriate integration measure.
The operational utility of $A_G(\varrho)$ is that it quantifies the extra work that is extractable from a quantum Szilard engine under a SSR when a system in the state $\varrho$ is used as a reference for the engine's working fluid.
In that case $k_B T A_G(\varrho)$, where $k_B$ is Boltzmann's constant and $T$ is the temperature of the thermal reservoir, is an achievable upper bound on the extra work \cite{VAWJarxiv,VAWJ}.
The asymmetry $A_G(\varrho)$ has a number of other important properties \cite{VAWJarxiv,VAWJ}, but the salient one for us here is that it is non increasing under operations $\varrho\to\mathcal{O}[\varrho]$ that are $G$-covariant, i.e.
\beq
    A_G(\mathcal{O}[\varrho ])\le A_G(\varrho)\ ,
\eeq
where a $G$-covariant operation is one that satisfies $\mathcal{O}[T_g\varrho T^\dagger_g]=T_g\mathcal{O}[\varrho]T^\dagger_g$ for all $g\in G$.

In particular, the U(1) symmetry group
\beq   \label{U(1)}
        {\rm U(1)}=\left\{T(\phi)=\exp\left(-\ii {H_0 \over s}\phi\right): 0<\phi\le 2\pi\right\}\ ,
\eeq
is continuous and its corresponding twirl is given by
\beq
        {\cal G}_{\rm U(1)}[\varrho ]=\int_{2\pi}\frac{d\phi}{2\pi}T(\phi)\varrho  T^\dagger(\phi)
\eeq
where $H_0=s\sum_n n\ket{n}\bra{n}+s_0$ is the free Hamiltonian of the system, $s$ and $s_0$ represent an energy gap and ``vacuum'' energy parameters, respectively, and $\phi$ is a phase angle.
This symmetry represents the invariance to phase rotations and $A_{\rm U(1)}(\varrho )$ measures the phase coherence of $\varrho$ in terms of how $\varrho$ breaks the U(1) symmetry.
The U(1)-covariant operations $\mathcal{O}[\cdot]$ satisfy
\beq  \label{U(1) covariant operation}
    \hspace{-5mm}\mathcal{O}\left[\exp\left(-\ii {H_0 \over s}\phi\right)\varrho \exp\left(\ii {{H_0}\over s}\phi \right)\right]=\exp\left(-\ii {H_0\over s}\phi\right)\mathcal{O}[\varrho]\exp\left(\ii {H_0\over s}\phi\right)
\eeq
for all values of $\phi$ in a $2\pi$ interval.  In other words, U(1)-covariant operations commute with the phase-shifting operation.  If we apply this to the reservoir state $\ket{\eta_{L,l_0}}$ then we find
\begin{equation}
\mathcal{G}_G[\ket{\eta}\bra{\eta}] = \frac{\id}{L} ,
\end{equation}
so that the asymmetry is $A_G[\ket{\eta}\bra{\eta}] = \ln L$.

\begin{figure}[b!]  %%%%%%%%%%%%%%%%%%%%%%%%%%%%%%%%%%%%%
	\begin{center}
		\includegraphics[width=12.0cm]{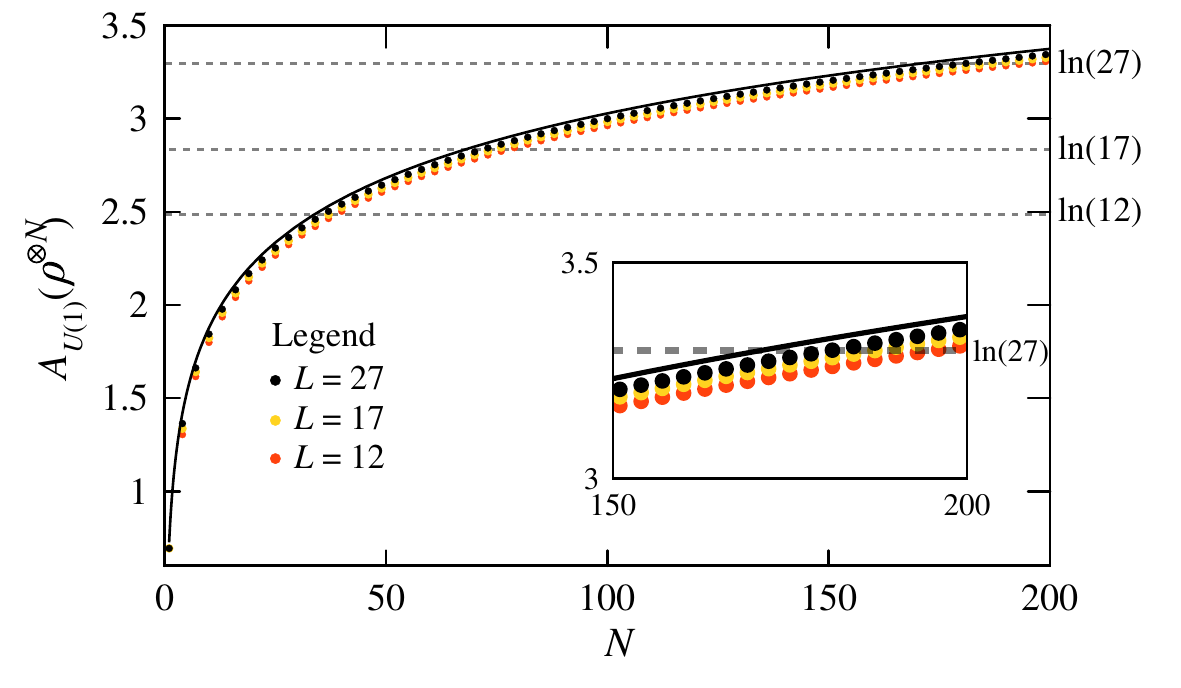}
	\end{center}
	\caption{Comparison of the exact values of $A_{U(1)}(\rho^{\otimes N})$ (discs) with the approximate values given by \eref{asymmetry of collection of systems} (continuous curve) for different values of $L$ and $N$. The red, yellow and black discs correspond to values of $A_{U(1)}(\rho^{\otimes N})$ for $L=12$, $17$ and $27$, respectively.  For clarity, the discs are plotted for every third value of $N$ starting from $N=1$.
Also plotted (as dotted grey lines) are the corresponding upper bounds on $ A_{\rm U(1)}({\rm Tr}_E[{\cal V}_N(\rho_0^{\otimes N}\otimes\sigma)])$.
The inset gives an enlarged view of the range $N=150$ to $200$.
		\label{fig:comparison}}
\end{figure}

The findings of CC, and \eref{same channel} in particular, suggest that the channel $\Phi_{\sigma,U}$ can produce an inexhaustible supply of systems in the state $\Phi_{\sigma,U}(\rho_0)$ and this has implications for the non increasing property of asymmetry.
To see this let the initial state of a collection of $N$ systems be $\rho_0^{\otimes N}$ where $\rho_0=\ket{\psi_0}\bra{\psi_0}$, $U$ be given by \eref{coherent operation} and the reservoir initially be in the state $\sigma=\ket{\eta_{L,l_0}}\bra{\eta_{L,l_0}}$ given by \eref{reservoir}.
This yields $\sigma'=\Lambda_{\rho_0,U}(\sigma)$, $\sigma''=\Lambda_{\rho_0,U}(\sigma')$ etc. and, using \eref{same channel}, we find that $\rho_0^{\otimes k}$ is transformed to
\beq
\fl        \cdots\otimes\Phi_{\sigma'',U}(\rho_0)\otimes\Phi_{\sigma',U}(\rho_0)\otimes\Phi_{\sigma,U}(\rho_0)
        &=\cdots\otimes\Phi_{\sigma,U}(\rho_0)\otimes\Phi_{\sigma,U}(\rho_0)\otimes\Phi_{\sigma,U}(\rho_0)\\
        &=\cdots\otimes\rho\otimes\rho\otimes\rho
\eeq
where $\rho$ is given by \eref{Eq28} with $\theta=0$, i.e.
\beq    \label{channel output state}
        \rho=\frac{1}{2}\left[\ket{\psi_0}\bra{\psi_0}+\ket{\psi_1}\bra{\psi_1}
        +\left(1-\frac{1}{L}\right)(\ket{\psi_0}\bra{\psi_1}+\ket{\psi_1}\bra{\psi_0})\right]\ .
\eeq
In \ref{app:asymmetry of rho N} we show that the asymmetry of the collection of systems is given approximately by
\beq    \label{asymmetry of collection of systems}
        A_{\rm U(1)}(\rho^{\otimes N})\approx \frac{1}{2}\ln\left(\frac{N\pi e}{2}\right)
\eeq
for large $L$ in the limit of large $N$.
\Fref{fig:comparison} shows that \eref{asymmetry of collection of systems} is a good approximation even for relatively small values of $L$ and $N$.
The fact that the right side of \eref{asymmetry of collection of systems} diverges as $N$ tends to infinity implies that the reservoir can be used to generate a collection of systems in a state that has \emph{unbounded} coherence.
Yet this conflicts with the physical requirement that the asymmetry must be non-increasing under physical operations.
Once again, the resolution lies in the correlations between the qubits that are omitted in the simple channel picture.
It is clear that these correlations are a fundamental component of the final multi-qubit state.

    %%%%%%%%%%%%%%%%%%%%%%%%%%%%%%%%%%%%%

\section{Discussion and Conclusion}

The validity of the key equations of CC, reproduced here as \eref{approx U rho U}, \eref{invariance of Delta}
and \eref{same channel}, is not in dispute.  These equations imply that each system $S_i$, if considered on its own (i.e.
in the absence of information about the state of any other system $S_{j\ne i}$), will have a reduced density operator given by $\rho$ in \eref{channel output state}.
The fact that the reduced density operator is $\rho$---regardless of how many prior times the reservoir has been used to prepare other systems---may appear to be extraordinary.
This situation simply reflects, however, the invariance of the single-system reduced density operator to the order in which the systems are prepared.
This invariance is apparent in the commutativity of the operators $V_i(U)$ defined according to \eref{defn of V(U)} for different systems $S_i$.
For example, it is straightforward to see that $V_1(U)V_2(U)= V_2(U)V_1(U)$ and it follows that this commutability property generalises to any two systems $S_i$ and $S_j$.
This leads to a crucial point: the dynamics of the interaction between the reservoir and the systems are \emph{invariant with respect to the ordering of the preparation of the systems}.

\begin{figure}[b!]  %%%%%%%%%%%%%%%%%%%%%%%%%%%%%%%%%%%%%
	\begin{center}
		\includegraphics[width=12.0cm]{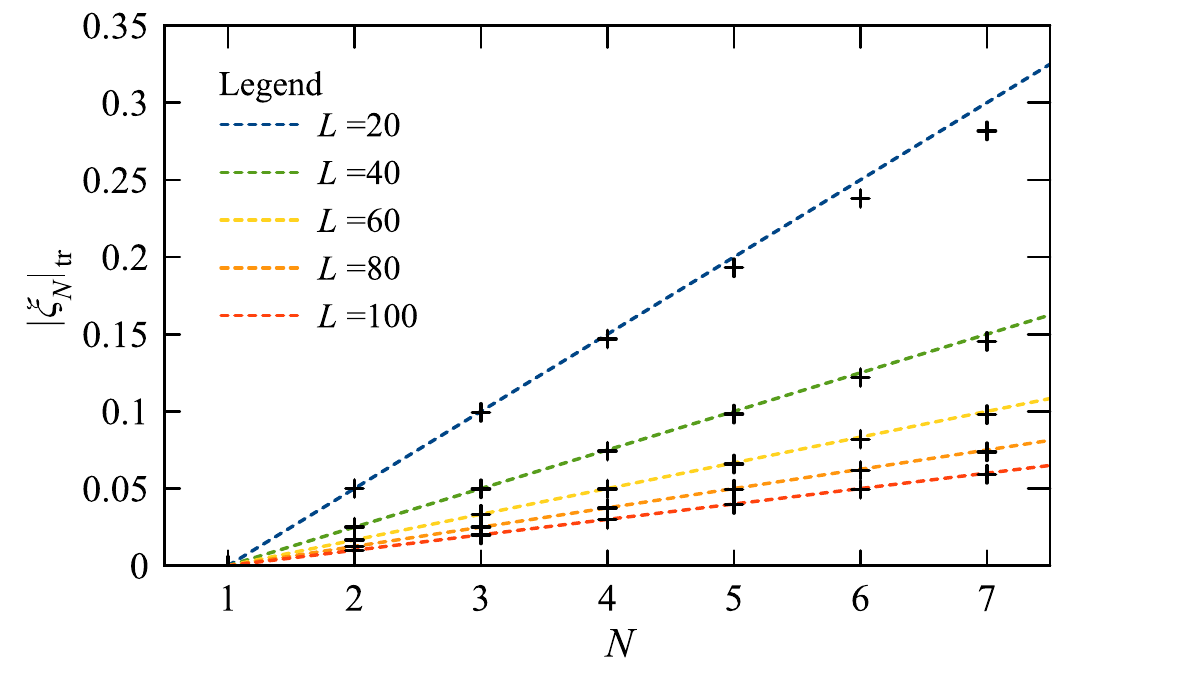}
	\end{center}
	\caption{The trace norm of the repeatability error $\xi_N$ as a function of the number of systems $N$ for various values of $L$. The crosses represent exact values of $|\xi_N|_{\mbox{tr}}$ and the dashed lines represent the approximation $|\xi_N|_{\mbox{tr}}\approx (N-1)/L$ derived in \ref{app:Repeatability error} for $1\le N\ll L$.
		\label{fig:error}}
\end{figure}    %%%%%%%%%%%%%%%%%%%%%%%%%%%%%%%%%%%%%

The invariance is the reason that every system prepared using the same reservoir, if considered on its own, has the same reduced density operator, $\rho$.
It does not, however, imply that the preparation of the systems is catalytic or even repeatable, as claimed in CC.
Rather, it merely implies that if \emph{one} system is examined, it will be found to be in the state $\rho$ regardless of the order in which it is prepared.
If, instead, \emph{two} systems are examined, they will be found in the state $\tr_E[{\cal V}_2(\rho_0^{\otimes 2}\otimes\sigma)]$ regardless of the order in which they are prepared.
To determine whether the preparation of a system is repeatable in the sense that another system is able to be prepared in the same state as the first, we need to compare the actual prepared state of \emph{both} systems in question, i.e.
$\tr_E[{\cal V}_2(\rho_0^{\otimes 2}\otimes\sigma)]$, with the state that represents both systems being prepared in the same state, i.e.
$\rho^{\otimes 2}$.
The fact that the state of two processed qubits is not $\rho^{\otimes 2}$ is a direct demonstration that the preparation is not repeatable.
In general, the \emph{repeatability error} in the preparation of $N$ systems is given by the difference $\xi_N=\tr_E[{\cal V}_N(\rho_0^{\otimes N}\otimes\sigma)]-\rho^{\otimes N}$. \Fref{fig:error} shows how the trace norm of $\xi_N$ grows linearly with $N$  for $1\le N\ll L$.
Given that it is the neglect of this error that leads to the paradoxical results discussed in preceding sections, it follows that the non-repeatability of the preparation cannot be ignored or even eliminated in principle---rather the non-repeatability of the preparation stands as a  necessity for consistency with basic quantum principles.
In conclusion, we can say, quite categorically, that \emph{coherence is not catalytic}.

\ack

The authors thank I.~Adagideli, D.~Jennings, T.~Rudolph and J.~\AA{}berg for helpful discussions.  J.A.V. thanks the Australian Research Council (LP140100797) and the Lockheed Martin Corporation for financial support, and acknowledges discussions with S.~Bedkihal.  S.M.B. thanks the Royal Society for support (RP150122).

\appendix

\section{Asymmetry of $\rho^{\otimes N}$\label{app:asymmetry of rho N}}

To derive a closed expression for the asymmetry
\beq
        \label{A_G(rho^N)}
        A_{\rm U(1)}(\rho^{\otimes N}) &= S(\mathcal{G}_{\rm U(1)}[\rho^{\otimes N}])-S(\rho^{\otimes N})\ ,
\eeq
where $\rho$ is given by \eref{channel output state} in the main text, we first deduce a number of preliminary results, as follows.
In places we treat the energy eigenstates $\ket{\psi_0}$ and $\ket{\psi_1}$ as the eigenstates of the $z$ component of angular momentum of a fictitious spin-1/2 particle with corresponding eigenvalues $-\hbar/2$ and $\hbar/2$.
This allows us to use the Dicke state basis $\{\ket{J,M;\boldsymbol{\lambda},i}\}$ where $J$ and $M$ are the analogous angular momentum quantum numbers and $\boldsymbol{\lambda}\equiv(\lambda_1,\lambda_2)$ and $i$ are quantum numbers that label different permutations of the systems \cite{Arecchi,BartlettWiseman}.
The quantum numbers satisfy $0\le J \le N/2$, $-J\le M \le J$, $N=\lambda_1+\lambda_2$, $\lambda_1\ge\lambda_2$, $2J=\lambda_1-\lambda_2$, $1\le i \le {}^NC_{\lambda_2}-{}^NC_{\lambda_2-1}$ \cite{Arecchi}.
They are all integer valued if the number of systems, $N$, is even. For brevity, we limit the following discussion to just this case; the extension to odd values of $N$ is, however, straightforward.
A $\rm U(1)$ phase rotation in the energy basis is equivalent to a spatial rotation about the $z$ axis in the Dicke basis.
As rotations leave the subspace $\{\ket{J,M;\boldsymbol{\lambda},i}: -J\le M\le J\}$ invariant, it is useful to express the Dicke states using the notation of a tensor product $\ket{J,M;\boldsymbol{\lambda},i}=\ket{J,M}\otimes\ket{\boldsymbol{\lambda},i}$ because then rotations have the form $R\otimes \id$, where $R$ is a $\rm SU(2)$ rotation operator that operates on the $\ket{J,M}$ component and $\id$ is the identity operator that operates on the $\ket{\boldsymbol{\lambda},i}$ component \cite{BartlettWiseman}.
With this notation, the $\rm U(1)$ twirl operation on $\rho^{\otimes N}$  is represented by
\beq
        \label{U(1) twirl as integral}
        {\cal G}_{\rm U(1)}[\rho^{\otimes N}]=\int_{2\pi}\frac{\dd\theta}{2\pi}\left(e^{i\theta J_z}\otimes \id\right) \rho^{\otimes N}  \left(e^{-i\theta J_z}\otimes \id\right)\ .
\eeq
Here and in the following, $J_\mu=\sum_n \sigma_\mu^{(n)}$ for $\mu=x$, $y$ or $z$ are components of the total angular momentum operator for the collection of systems and $\sigma_\mu^{(n)}$ are the corresponding Pauli spin operators for the $n$th spin-$1/2$ system.
As the twirl is a linear operation, we can separate its effect on individual terms in the Dicke-state expansion of density operator.
In particular, terms proportional to
\beq
        \label{terms in Dicke basis}
        \op{J,M}{J',M'}
        \otimes\op{\boldsymbol{\lambda},i}{\boldsymbol{\lambda}',i'}
\eeq
are reduced to zero by the twirl if $M\ne M'$ and left unchanged otherwise.
It follows that an equivalent form of the twirl operation is given by
\beq
        \label{U(1) twirl as projections}
        {\cal G}_{\rm U(1)}[\rho^{\otimes N}]=\sum_{M=-N/2}^{N/2}\Pi_M\rho^{\otimes N}\Pi_M
\eeq
where
\beq
        \label{projection op M Dicke basis}
        \Pi_M=
        \sum_{J=|M|}^{N/2} \left(\op{J,M}{J,M}
        \otimes\sum_{i=1}^{\Gamma_J}
            \op{\boldsymbol{\lambda},i}{\boldsymbol{\lambda},i}\right)
\eeq
is a projection operator that projects onto the eigenspace of $J_Z$ associated with eigenvalue $M$,
\beq
        \Gamma_J=\cases{
                        1&  for $|J|=N/2$\\
                        \mbox{${N \choose {\lambda_2}}-{N \choose {\lambda_2-1}}$}  & otherwise }
\eeq
and $\lambda_2=N/2-J$.
It is straightforward to show that the right side of \eref{U(1) twirl as projections} has the same effect on the terms in \eref{terms in Dicke basis} as the right side of \eref{U(1) twirl as integral}.
An equivalent form of $\Pi_m$ is given in the energy basis by
\beq
        \label{projection op M energy basis}
        \Pi_M=
        \sum_{z=0}^{2^N-1}\delta_{h(\tilde{z}),N/2+M}\op{\tilde{z}}{\tilde{z}}
\eeq
where $\ket{\tilde{z}}\equiv\bigotimes_{n=1}^N \ket{\psi_{\tilde{z}_n}}$ represents the collective state of the $N$ systems in the $\ket{\psi_0}$, $\ket{\psi_1}$ basis, $\tilde{z}$ is a binary representation of $z$, $\tilde{z}_n$ is the $n$th bit of $\tilde{z}$, and $h(\tilde{z})$ is the Hamming weight of $\tilde{z}$ (i.e. the number of 1's in $\tilde{z}$).

Let the projection of $\rho^{\otimes N}$ be represented by
\beq
        \label{defn Q_M}
        \mathcal{Q}_M=\frac{1}{p_M}\Pi_M \rho^{\otimes N}\Pi_M
\eeq
where $\mathcal{Q}_M$ is a normalised density operator and $p_M$ is the normalisation constant.  The value of $p_M$ can be calculated in the energy basis as follows. We reexpress $\rho$ from \eref{channel output state} as
\beq
        \rho = \frac{1}{2}\left( \id_s + a\sigma_x\right)
\eeq
where $a=1-1/L$, $\id_s=\op{\psi_0}{\psi_0}+\op{\psi_1}{\psi_1}$ is a system identity operator, and $\sigma_x=\op{\psi_0}{\psi_1}+\op{\psi_1}{\psi_0}$ and make use of \eref{projection op M energy basis} to arrive at
\beq
    \label{example proj M}
        \Pi_M\rho^{\otimes N}\Pi_M &=
                \sum_{z=0}^{2^N-1}\delta_{h(\tilde{z}),\ell}\op{\tilde{z}}{\tilde{z}}
                \frac{1}{2^N}\left( \id_s + a\sigma_x\right)^{\otimes N}
                \sum_{z'=0}^{2^N-1}\delta_{h(\tilde{z}'),\ell}\op{\tilde{z}'}{\tilde{z}'}\\
                  &=\frac{1}{2^N}\sum_{z,z'=0}^{2^N-1}\delta_{h(\tilde{z}),\ell}\delta_{h(\tilde{z}'),\ell}
                  a^{h(\tilde{z}\oplus\tilde{z}')}\op{\tilde{z}}{\tilde{z}'}
    \label{example proj M 2}
\eeq
where $\ell=N/2+M$ and $A\oplus B$ represents the bitwise exclusive-or operation on the binary numbers $A$ and $B$.  The last result was derived by noting three things: (i) each $\sigma_x$ operator in \eref{example proj M} induces a bit flip at a unique location in the label of the state $\ket{\tilde{z}'}$, (ii) only one term in the expansion of the product in $\bra{\tilde{z}}(\id + a\sigma_x)^{\otimes N}\ket{\tilde{z}}$ is nonzero for $h(\tilde{z})=h(\tilde{z}')=\ell$, and (iii) the number of bit flips to make $\tilde{z}'$ equal to $\tilde{z}$ (i.e. the Hamming distance between $\tilde{z}'$ and $\tilde{z}$) gives the power of $a$ in the nonzero term in (ii).  Taking the trace of \eref{example proj M 2} then yields
\beq
p_M &=\tr (\Pi_M\rho^{\otimes N}\Pi_M)
        =\frac{1}{2^N}\sum_{z=0}^{2^N-1}\delta_{h(\tilde{z}),N/2+M}\\
        &=\frac{1}{2^N}{N \choose {N/2+M}}\ .
    \label{p_M}
\eeq

Next, we find the representation of $\rho^{\otimes N}$ in the Dicke basis by first diagonalising $\rho$:
\beq
        \rho=a\op{1}{1}+b\op{0}{0}
\eeq
where $a=1-1/L$ as above, $b=1/L$,  $\ket{0}\equiv(\ket{\psi_0}-\ket{\psi_1})/\sqrt{2}$ and $\ket{1}\equiv(\ket{\psi_0}+\ket{\psi_1})/\sqrt{2}$.  The tensor product $\rho^{\otimes N}$ has a simple binomial expansion in this basis, i.e.
\beq
        \label{rho^N in Dicke basis}
        \rho^{\otimes N}=\sum_{k=0}^N {N \choose k} a^{N-k}b^k\mathcal{R}_k
\eeq
where  $\mathcal{R}_k$ is the normalised density operator
\beq
        \label{R_k}
        \mathcal{R}_k=\frac{1}{{N \choose k}}\sum_{x=0}^{2^N-1}\delta_{h(\tilde{x}),k}\op{\tilde{x}}{\tilde{x}}\ .
\eeq
Here $\ket{\tilde{x}}\equiv\bigotimes_{n=1}^N \ket{\tilde{x}_n}$ represents the collective state in the $\ket{0}$, $\ket{1}$ basis, $\tilde{x}$ is a binary representation of $x$, and $\tilde{x}_n$ is the $n$th bit of $\tilde{x}$.
The sum in \eref{R_k} would be equal to the sum in \eref{projection op M energy basis} for $M=k-N/2$ if the states $\ket{0}$ and $\ket{1}$ in \eref{R_k} were replaced with $\ket{\psi_0}$ and $\ket{\psi_1}$, respectively.
As $\ket{0}$ and $\ket{1}$ are related to $\ket{\psi_0}$ and $\ket{\psi_1}$ by a rotation of $\pi/2$ around the $y$ axis, i.e. $\ket{0}=e^{i\sigma_y\pi/2}\ket{\psi_0}$ and $\ket{1}=e^{i\sigma_y\pi/2}\ket{\psi_1}$, it follows that
\beq
        \label{R_k in Dicke basis}
        \mathcal{R}_k=\frac{1}{{N \choose k}}
        \left(e^{iJ_y\pi/2}\otimes \id\right)\Pi_{k-N/2} \left(e^{-iJ_y\pi/2}\otimes \id \right)\ .
\eeq

We now use the last result to express $\mathcal{Q}_M$ in \eref{defn Q_M} in the Dicke basis.  Substituting for $\rho^{\otimes N}$ in \eref{defn Q_M} using \eref{rho^N in Dicke basis} and \eref{R_k in Dicke basis}, i.e.
\beq
        \mathcal{Q}_M=\frac{1}{p_M}\sum_{k=0}^{N}a^{N-k}b^k\,\Pi_M \left(e^{iJ_y\pi/2}\otimes \id\right)\Pi_{k-N/2} \left(e^{-iJ_y\pi/2}\otimes \id \right)\Pi_M\ ,
\eeq
replacing $\Pi_M$ using \eref{projection op M Dicke basis} and then using the fact that rotations leave the value of $J$ unchanged yields
\beq
 \fl       \label{projected state Q_M diagonal}
        \mathcal{Q}_M=\frac{1}{p_M}\sum_{k=0}^{N}
            \sum_{J=J_0}^{N/2} a^{N-k}b^k\,|d^J_{M,k-N/2}(\pi/2)|^2\op{J,M}{J,M}\otimes\sum_{i=1}^{\Gamma_J}
            \op{\boldsymbol{\lambda},i}{\boldsymbol{\lambda},i}
\eeq
where
\beq
        \label{J0 max M, k-N/2}
        J_0\equiv\max\{|M|,|k-N/2|\}
\eeq
and $d^J_{M',M}(\beta)=\ip{J,M'|e^{-iJ_y\beta}|J,M}$ are the matrix elements of the rotation operator $e^{-iJ_y\beta}$ \cite{Rose}.
Conveniently, \eref{projected state Q_M diagonal} gives the diagonal representation of $\mathcal{Q}_M$.

The projected state operator $\mathcal{Q}_M$ is normalised and so taking the trace of \eref{projected state Q_M diagonal} and substituting for $p_M$ using \eref{p_M} yields
\beq
        1=\sum_{k=0}^{N} a^{N-k}b^k\left(\sum_{J=J_0}^{N/2}|d^J_{M,k-N/2}(\pi/2)|^2 \frac{2^N}{{N \choose{N/2+M}}}\Gamma_J\right)\ .
\eeq
The fact that this holds for all positive values of $a$ with $b=1-a$ implies that the expression in the large brackets is equal to ${N \choose k}$.  To see this, treat the right side of the equation
\beq
        1= \sum_{k=0}^{N}a^{N-k}(1-a)^k x_k
\eeq
as a polynomial in $a$ and solve for $x_k$. For example, collecting powers of $a$,
\beq
        1= \sum_{r=0}^{N}a^{r}
        \sum_{s=0}^{r}{N \choose {s}}(-1)^{s} x_{s+N-r}\ ,
\eeq
and equating coefficients of like powers of $a$ on both sides yields $x_N = 1$ for $a^0$, $x_{N-1} = N$ for $a^1$, $x_{N-2} = {N \choose 2}$ for $a^2$ and so on, with the general solution being $x_k={N \choose k}$. Thus, we find the useful result that
\beq
        \label{sum of d}
        \sum_{J=J_0}^{N/2}|d^J_{M,k-N/2}(\pi/2)|^2\Gamma_J
        =\frac{1}{2^N}{N \choose{N/2+M}}{N \choose k}\ .
\eeq

The von Neumann entropy $S(\mathcal{Q}_M)$ follows directly from the diagonal representation of $\mathcal{Q}_M$ given in \eref{projected state Q_M diagonal}, i.e.
\beq
\fl
  S(\mathcal{Q}_M) &=-\sum_{k=0}^{N}
            \sum_{J=J_0}^{N/2} \sum_{i=1}^{\Gamma_J} a^{N-k}b^k\,\frac{|d^J_{M,k-N/2}(\pi/2)|^2}{p_M} \ln\left(a^{N-k}b^k\,\frac{|d^J_{M,k-N/2}(\pi/2)|^2}{p_M}\right)\ .
\eeq
Performing the sum over $i$, substituting for $p_M$ using \eref{p_M} and reexpressing the logarithm, i.e.
\beq
\fl
        S(\mathcal{Q}_M)    =-\sum_{k=0}^{N}
            \sum_{J=J_0}^{N/2} \Gamma_J a^{N-k}b^k\,\frac{|d^J_{M,k-N/2}(\pi/2)|^2 2^N}{{N \choose {N/2+M}}}\left[\ln\left(a^{N-k}b^k\right)+\ln\left(\frac{|d^J_{M,k-N/2}(\pi/2)|^2 2^N}{{N \choose {N/2+M}}}\right)\right]\ ,\nonumber\\
              \label{S(Q_M) intermediate}
\eeq
and then using \eref{sum of d} yields
\beq
        \label{S(Q_M)=binomial sum + Delta S}
        S(\mathcal{Q}_M)    &=-\sum_{k=0}^{N}
             a^{N-k}b^k{N \choose k}\ln\left(a^{N-k}b^k\right) +\epsilon_M
\eeq
where
\beq
\fl        \label{epsilon_M}
    \epsilon_M=-\sum_{k=0}^{N} \sum_{J=J_0}^{N/2} \Gamma_J a^{N-k}b^k\,\frac{|d^J_{M,k-N/2}(\pi/2)|^2 2^N}{{N \choose {N/2+M}}}\ln\left(\frac{|d^J_{M,k-N/2}(\pi/2)|^2 2^N}{{N \choose {N/2+M}}}\right)\ .
\eeq
Noting that the binomial coefficient ${N \choose k}$ in \eref{S(Q_M)=binomial sum + Delta S} represents the number of equal-likely events with probability $a^{N-k}b^k$, we recognise the first term as being equal to $S(\rho^{\otimes N})$, i.e.
\beq
        \label{S(Q_M)=S(rho N) + Delta S_M}
        S(\mathcal{Q}_M) = S(\rho^{\otimes N})+\epsilon_M\ .
\eeq
Next we derive an approximate expression for $\epsilon_M$ that is valid for large $L$ (i.e. for $a\approx 1$ and $b\approx 0$) in the limit that $N\to\infty$ using the facts that (i) the projected state $\mathcal{Q}_M$ is distributed binomially according to $p_M$ in \eref{p_M}, and (ii) from \eref{sum of d} the sum \beq
        \sum_{J=J_0}^{N/2}\Gamma_J a^{N-k}b^k\,\frac{|d^J_{M,k-N/2}(\pi/2)|^2 2^N}{{N \choose {N/2+M}}}=a^{N-k}b^k{N \choose k}
\eeq
is a binomial distribution over $k$ centred on $k\approx b N$.
According to (i), it is only the projected states $\mathcal{Q}_M$ with $M\approx 0$ to order $\sqrt{N}$ that contribute significantly in \eref{twirl} and so we limit our attention to $M\approx 0$.
In regards to (ii), in \eref{epsilon_M} the terms that contribute significantly to the sum over $k$ are those for which $k\approx b N$ to order $\sqrt{N}$, and so ignoring all other terms means that $J_0=N(1/2-b)$ according to \eref{J0 max M, k-N/2}, and so we only need to consider terms in the sum over $J$ in the range $J=N(1/2-b),\ldots,N/2$.
These terms, with $M=0$ and $k=b N$, have the form
\beq
        \label{terms}
       \Gamma_{ N(1/2-c)}\frac{|d^{N(1/2-c)}_{0,-N(1/2-b)}(\pi/2)|^2 2^N}{{N \choose {N/2}}}\ln\left(\frac{|d^{N(1/2-c)}_{0,-N(1/2-b)}(\pi/2)|^2 2^N}{{N \choose {N/2}}}\right)
\eeq
where $0<c\le b \ll 1/2$.
The Wigner-d matrix elements have the form \cite{Arecchi,Rose}
\beq
        d^{j}_{0,m}(\pi/2)=\sum_{n} \frac{(-1)^n}{n!(n-m)!}\frac{(j+n)!}{(j-n)!}\left(\frac{(j+m)!}{(j-m)!}\right)^{1/2}
        \frac{1}{2^n}
\eeq
where sum is over values of $n$ which give non-negative values for the arguments of the factorials, and thus is from $0$ to $N(1/2-c)$. Substituting $j=N(1/2-c)$ and $m=-N(1/2-b)$, i.e.
\beq
 \fl       d^{N(1/2-c)}_{0,-N(1/2-b}(\pi/2)
            &\approx\sum_{n=0}^{N(1/2-c)} \frac{(-1)^n}{n![n+N(1/2-b)]!}\frac{[N(1/2-c)+n]!}{[N(1/2-c)-n]!}
        \left\{\frac{[N(b-c)]!}{[N(1-b-c)]!}\right\}^{1/2}\frac{1}{2^n}\ , \nonumber
\eeq
and making the approximations $N(1/2-c)\approx N(1/2-b)\approx N/2$ in the large $N$ limit gives
\beq
        d^{N(1/2-c)}_{0,-N(1/2-b}(\pi/2)
        &\approx \left\{\frac{[N(b-c)]!}{N!}\right\}^{1/2}\frac{1}{(N/2)!}\sum_{n=0}^{N/2}{N/2 \choose n}\left(\frac{-1}{2}\right)^n\\
        &=\left\{\frac{[N(b-c)]!}{N!}\right\}^{1/2}\frac{1}{(N/2)!2^{N/2}}
\eeq
and so
\beq
        \frac{|d^{N(1/2-c)}_{0,-N(1/2-b)}(\pi/2)|^2 2^N}{{N \choose {N/2}}}\to 0
\eeq
as $N\to\infty$.  Correspondingly, the terms in \eref{terms} vanish in the same limit and so we find from \eref{epsilon_M} with $M=0$ that
\beq
        \label{epsilon_M to 0}
        \epsilon_0 \to 0 \mbox{ as }N\to\infty\ .
\eeq

The stage is finally set for deriving an expression for $A_{\rm U(1)}(\rho^{\otimes N})$.
From \eref{U(1) twirl as projections} and \eref{defn Q_M} we find
\beq
        \label{G(rho^N) = sum p_M Q_M}
        \mathcal{G}_{\rm U(1)}[\rho^{\otimes N}]
        =\sum_{M=-N/2}^{N/2}\Pi_M \rho^{\otimes N}\Pi_M
        =\sum_{M=-N/2}^{N/2}p_M\mathcal{Q}_M
\eeq
and using the diagonal representation of $\mathcal{Q}_M$ in \eref{projected state Q_M diagonal} gives
\beq
\fl        S(\mathcal{G}_{\rm U(1)}[\rho^{\otimes N}])
        &=-\sum_{M=-N/2}^{N/2}\sum_{k=0}^{N}
            \sum_{J=J_0}^{N/2}\sum_{i=1}^{\Gamma_J} a^{N-k}b^k\,|d^J_{M,k-N/2}(\pi/2)|^2 \ln\left(a^{N-k}b^k\,|d^J_{M,k-N/2}(\pi/2)|^2\right)\nonumber\\
        &=-\sum_{M=-N/2}^{N/2}\sum_{k=0}^{N}
            \sum_{J=J_0}^{N/2}\Gamma_J a^{N-k}b^k\,|d^J_{M,k-N/2}(\pi/2)|^2 \ln\left(a^{N-k}b^k\,|d^J_{M,k-N/2}(\pi/2)|^2\right)\ .\nonumber\\
        \label{S(G(rho^N)) exact}
\eeq
Next, using \eref{rho^N in Dicke basis} and \eref{R_k in Dicke basis} we find
\beq
\fl        S(\rho^{\otimes N})=S\left[\sum_{k=0}^N {N \choose k} a^{N-k}b^k\frac{1}{{N \choose k}}
        \left(e^{iJ_y\pi/2}\otimes \id\right)\Pi_{k-N/2} \left(e^{-iJ_y\pi/2}\otimes \id \right)\right]
\eeq
which reduces to
\beq
        S(\rho^{\otimes N})=S\left(\sum_{k=0}^N  a^{N-k}b^k\Pi_{k-N/2}\right)
\eeq
because the rotation about the $y$ axis does not change the entropy.  According to the representations in \eref{projection op M Dicke basis}, \eref{projection op M energy basis} or \eref{R_k in Dicke basis}, the projection operator $\Pi_{k-N/2}$ projects onto a subspace of dimension ${N \choose k}$ and so
\beq
        S(\rho^{\otimes N})
        =-\sum_{k=0}^{N}
             a^{N-k}b^k{N \choose k} \left[\ln\left(a^{N-k}b^k\right)\right]\ .
\eeq
Multiplying by unity in the form of $\sum_{M=-N/2}^{N/2}\frac{1}{2^N}{N \choose{N/2+M}}=1$ and then using \eref{sum of d} we find
\beq
 \fl       S(\rho^{\otimes N})
        &=-\sum_{M=-N/2}^{N/2}\sum_{k=0}^{N}
             a^{N-k}b^k\frac{1}{2^N}{N \choose{N/2+M}}{N \choose k} \left[\ln\left(a^{N-k}b^k\right)\right]\nonumber\\
        &=-\sum_{M=-N/2}^{N/2}\sum_{k=0}^{N}
            \sum_{J=J_0}^{N/2} \Gamma_J a^{N-k}b^k\,|d^J_{M,k-N/2}(\pi/2)|^2 \left[\ln\left(a^{N-k}b^k\right)\right]
        \label{S(rho^N) exact}
\eeq
and so substituting for $S(\mathcal{G}_{\rm U(1)}[\rho^{\otimes N}])$ and $S(\rho^{\otimes N})$ in \eref{A_G(rho^N)} using \eref{S(G(rho^N)) exact} and \eref{S(rho^N) exact} finally gives an exact expression for $A_{\rm U(1)}(\rho^{\otimes N})$ as
\beq
\fl        \label{A_U(rho^N) exact}
        A_{\rm U(1)}(\rho^{\otimes N})
        &=-\sum_{M=-N/2}^{N/2}\sum_{k=0}^{N}
            \sum_{J=J_0}^{N/2} a^{N-k}b^k\,\Gamma_J |d^J_{M,k-N/2}(\pi/2)|^2 \ln\left(|d^J_{M,k-N/2}(\pi/2)|^2\right)\ .
\eeq

More useful, however, is an approximate expression that is valid for large $L$ (i.e. for $a\approx 1$ and $b\approx 0$) in the limit that $N\to\infty$.  To derive it, note that the projection operator $\Pi_M$ defined in \eref{projection op M Dicke basis} projects onto disjoint subspaces for different values of $M$, and so the projections $\mathcal{Q}_M$ form a set of mutually orthogonal density operators.
Making use of this together with \eref{G(rho^N) = sum p_M Q_M} gives
\beq
        \label{S(G)=H + sum p_m S(Q_M)}
        S(\mathcal{G}_{\rm U(1)}[\rho^{\otimes N}])=H(\{p_M\}) + \sum_{M=-N/2}^{N/2}p_M S(\mathcal{Q}_M)
\eeq
where $H(\{p_M\})=-\sum_{M=-N/2}^{N/2}p_M\ln(p_M)$ is the Shannon entropy associated with the set of probabilities $\{p_M\}$. Substituting into \eref{A_G(rho^N)} and then recalling the results in \eref{S(Q_M)=S(rho N) + Delta S_M} and \eref{epsilon_M to 0} shows
\beq
        A_{\rm U(1)}(\rho^{\otimes N})\to
        H(\{p_M\}) \mbox{ as } N\to \infty\ .
\eeq
Using the Gaussian approximation to the binomial distribution further simplifies the result to
\beq
        \label{A_U(rho^N)=ln(N)}
        A_{\rm U(1)}(\rho^{\otimes N})\approx\frac{1}{2}\ln(\frac{N\pi e}{2})
\eeq
which appears as \eref{asymmetry of collection of systems} in the main text.

\section{Repeatability error \label{app:Repeatability error}}

The \emph{repeatability error} $\xi_N$ in the preparation of $N$ systems is defined by
\beq
        \label{error E}
     \xi_N= \tr_E[{\cal V}_N(\rho_0^{\otimes N}\otimes\sigma)] - \rho^{\otimes N}\ .
\eeq
Using $U\ket{\psi_{0}}=(\ket{\psi_{0}}+\ket{\psi_{1}})/\sqrt{2}$ for $U$ in  \eref{defn of V(U)} and $\rho_0=\op{\psi_{0}}{\psi_{0}}$ gives
\beqn
\fl     \tr_E[{\cal V}_N(\rho_0^{\otimes N}\otimes\sigma)]
     = {1\over 2^N}\sum_{n,n',m,m',\ldots}
        \op{\psi_n}{\psi_{n'}}\otimes\op{\psi_m}{\psi_{m'}}\otimes\ldots \tr_E(\Delta^{-n-m-\ldots}\sigma\Delta^{n'+m'+\ldots})
\eeqn
and evaluating the partial trace over the reservoir yields
\beqn
\fl     \tr_E[{\cal V}_N(\rho_0^{\otimes N}\otimes\sigma)]
     = {1\over 2^N}\sum_{n,n',m,m',\ldots}
        \op{\psi_n,\psi_m\ldots}{\psi_{n'},\psi_{m'}\ldots}
        \left(1-\mbox{$\frac{|n-n'+m-m'+\ldots|^{(L)}}{L}$}\right)\ ,
\eeqn
where $|n|^{(L)}\equiv\min\{L,|n|\}$ and $\ket{\psi_n,\psi_m\ldots}\equiv\ket{\psi_n}\otimes\ket{\psi_m}\otimes\ldots$.
Similarly, we find
\beqn
     \rho^{\otimes N}
     = {1\over 2^N}\sum_{n,n',m,m',\ldots}
        \op{\psi_n,\psi_m\ldots}{\psi_{n'},\psi_{m'}\ldots}
        \left(1-\mbox{$\frac{|n-n'|}{L}$}\right)\left(1-\mbox{$\frac{|m-m'|}{L}$}\right)\ldots\ ,
\eeqn
and so from \eref{error E} the repeatability error is given by
\beq
        \label{error_N exact}
     \xi_N&= {1\over 2^N}\sum_{n,n',m,m',\ldots}
        \op{\psi_n,\psi_m\ldots}{\psi_{n'},\psi_{m'}\ldots}\nonumber\\
        &\times \left[
        1-\mbox{$\frac{|n-n'+m-m'+\ldots|^{(L)}}{L}$}
        -\left(1-\mbox{$\frac{|n-n'|}{L}$}\right)\left(1-\mbox{$\frac{|m-m'|}{L}$}\right)\ldots\right]
        \ .
\eeq
An approximate expression for $\xi_N$ in the regime where $L\gg 1$ and $N\ll L$ is derived using the following four facts about the terms in \eref{error_N exact}: (i) $|n-n'+m-m'+\ldots|^{(L)}=|n-n'+m-m'+\ldots|$ for $N<L$, (ii) $(1-\frac{|n-n'|}{L})(1-\frac{|m-m'|}{L})\ldots=1-\frac{|n-n'|}{L}-\frac{|m-m'|}{L}-\ldots$ to first order in $1/L$ for $N\ll L$, (iii) the terms for which $|n-n'+m-m'+\ldots|\ll |n-n'|+|m-m'|+\ldots$ are far more abundant than the remaining terms for $1\ll N \ll L$, and (iv) the values of $|n-n'+m-m'+\ldots|$ are not necessarily negligible compared to those of $|n-n'|+|m-m'|+\ldots$ for values of $N$ of the order of unity.
The first three facts imply that the expression in square brackets in \eref{error_N exact} can be approximated by $\frac{1}{L}(|n-n'|+|m-m'|+\ldots)$ for $1\ll N \ll L$ whereas the fourth fact implies that this needs to be reduced to $\frac{1}{L}(|n-n'|+|m-m'|+\ldots - |n-n'+m-m'+\ldots|)$ to be useful for relatively small values of $N$.
Note that the expression $|n-n'+m-m'+\ldots|$ here is to be replaced with $|n-n'|$ for $N=1$ and that it contributes little for large $N$; this suggests an approximate expression that is valid for $N=1$ as well as $N\gg 1$ is given by $\frac{1}{L}(|n-n'|+|m-m'|+\ldots - |n-n'|)$ which is to be interpreted as zero for $N=1$ and $\frac{1}{L}(|m-m'|+\ldots)$ otherwise.
The corresponding approximate expression for the repeatability error is, therefore, $\xi_1=0$ for $N=1$ and
\beq
 \fl    \xi_N&\approx
        {1\over 2^N L}\sum_{n,n'}
        \op{\psi_n}{\psi_{n'}}\otimes\kern-1em
        \sum_{m,m',p,p'\ldots}
        \op{\psi_m,\psi_p\ldots}{\psi_{m'},\psi_{p'}\ldots}
                 \left(|m-m'|+|p-p'|+\ldots\right)\ ,\nonumber
\eeq
where there are $N-1$ terms in the bracketed expression, for $1 < N\ll L$.  The trace norm $|\xi_N|_{\mbox{tr}}=\tr(\sqrt{\xi_N^\dagger\xi_N})$ is then easily calculated to be
\beq
        |\xi_N|_{\mbox{tr}}\approx\frac{N-1}{L}\ .
\eeq
\Fref{fig:error} compares values given by this approximation with numerically calculated, exact values of $|\xi_N|_{\mbox{tr}}$ for a range of values of $L$.


\section*{References}

\begin{thebibliography}{99}
\bibitem{Aberg}
\AA{}berg J 2014 Catalytic coherence \textit{Phys. Rev. Lett.} {\bf 113} 150402 and  Supplementary Materials at
http://link.aps.org/supplemental/10.1103/PhysRevLett.113.150402.

\bibitem{Feynman}
Feynman R P, Leighton R B and Sands M 1965 \textit{The Feynman lectures on physics} vol. III
(Reading MA: Addison-Wesley)

\bibitem{Scovil}
Scovil H E D and Schultz-DuBois E O 1959 Three-level masers as heat engines \textit{Phys. Rev. Lett.}
{\bf 2} 262--263

\bibitem{Scully}
Scully M O 2017 Laser entropy arXiv:1708.06642

\bibitem{Jennings}
Korzekwa K, Lostaglio M, Oppenheim J and Jennings D 2016 {\it The extraction of work from quantum coherence} New J. Phys. {\bf 18}, 023045.

\bibitem{Malabarba}
Malabarba A S L, Short A J and Kammerlander P 2015 {\it Clock-driven quantum thermal engines} New J. Phys. {\bf 17}, 045027.

\bibitem{Marvian}
Marvian I and Lloyd S 2016 From clocks to cloners: Catalytic transformations under covariant operations and recoverability. arXiv: 1608.07325

\bibitem{Hayashi}
Hayashi M and Tajima H 2017 {\it Measurement-based formulation of quantum heat engines} Phys. Rev. A {\bf 95}, 032132.

\bibitem{Goold}
Goold J, Huber M, Riera A, del Rio L and Skrzypczyk P 2016 {\it The role of quantum information in thermodynamics---a topical review} J.Phys. A {\bf 49}, 143001.

\bibitem{Du}
Du S, Bai Z and Guo Y 2015 \emph{Conditions for coherence transformations under incoherent operations} Phys. Rev. A {\bf 91}, 052120. % http://dx.doi.org/10.1103/PhysRevA.91.052120 result (iii) on page 2 rules out Aberg's catalysis

\bibitem{Brando}
Brando F, Horodecki M, Ng N, Oppenheim J and Wehner S 2015 \emph{The second laws of quantum thermodynamics} P. Natl. Acad. Sci. USA {\bf 112}, 3275-3279. % http://dx.doi.org/10.1073/pnas.1411728112

\bibitem{Gallego}
Gallego R, Eisert J and  Wilming H 2016 \emph{Thermodynamic work from operational principles} New J. Phys. {\bf 18}, 103017. % http://dx.doi.org/http://stacks.iop.org/1367-2630/18/i=10/a=103017 treats genuine catalysis, but also treats approximate catalysis\end{thebibliography}

\bibitem{Ng2017}
Ng N H Y, Woods M P and Wehner S 2017 \emph{Surpassing the Carnot efficiency by extracting imperfect work} New J. Phys. {\bf 19}, 113005. % http://dx.doi.org/http://stacks.iop.org/1367-2630/19/i=11/a=113005

\bibitem{Ng2015}
Ng N H Y, Mancinska L, Cirstoiu C, Eisert J and Wehner S 2015 \emph{Limits to catalysis in quantum thermodynamics} New J. Phys. {\bf 17}, 085004. % http://dx.doi.org/http://stacks.iop.org/1367-2630/17/i=8/a=085004

\bibitem{BartlettNJP}
Bartlett S D, Rudolph T, Spekkens R W and Turner P S 2006 Degradation of a quantum reference frame \emph{New J. Phys.} {\bf 8}, 58. % http://dx.doi.org/10.1088/1367-2630/8/4/058 quant-ph/0602069

\bibitem{BartlettJMO}
Bartlett S D, Rudolph T, Sanders B C and Turner P S 2007 Degradation of a quantum directional reference frame as a random walk \emph{J. Mod. Optics} {\bf 54}, 2211-2221. % http://dx.doi.org/10.1080/09500340701289254

\bibitem{BartlettRMP}
Bartlett S D, Rudolph T and Spekkens R W 2007 Reference frames, superselection rules, and quantum information \emph{Rev. Mod. Phys.} {\bf 79}, 555--609. % http://dx.doi.org/10.1103/RevModPhys.79.555 arXiv:quant-ph 0610030

\bibitem{White}
White G A, Vaccaro J A and Wiseman H M 2009 The Consumption of Reference Resources \emph{AIP Conf. Proc.} {\bf 1110}, 79. % http://dx.doi.org/10.1063/1.3131381

\bibitem{Cover}
Cover T M and Thomas J A 1991 {\it Elements of information theory} (New York: Wiley)

\bibitem{Oxford}
Allen R E (ed.) 1984 {\it The Pocket Oxford Dictionary of Current English} 7th edn.
(Oxford: Clarendon Press)

\bibitem{Berzelius}
Berzelius J 1835 {\it {\AA}rsber\"{a}ttelse om framsteg i fysik och kemi}
(Stockholm: Royal Swedish Academy of Sciences) p. 245

\bibitem{KTH}
https://www.kth.se/en/che/archive/arkiv/berzelius-1.184145 accessed December 14 2017

\bibitem{Davy}
Davy H 1817 Some new experiments and observations on the combustion of gaseous mixtures,
with an account of a method of preserving a continued light in mixtures of inflammable gases and air without
flame \textit{Phil. Trans. R. Soc. Lond.} {\bf 107} 77--85

\bibitem{Chefles}
Chefles A 2000 Quantum state discrimination \textit{Contemp. Phys.} {\bf 41} 401--424.

\bibitem{SarahRev}
Barnett S M and Croke S 2009 Quantum state discrimination \textit{Adv. Opt. Photon.} {\bf 1} 238--278.

\bibitem{Helstrom}
Helstrom C W 1976 {\it Quantum Detection and Estimation Theory} (New York: Academic Press).

\bibitem{Stevebook}
Barnett S M 2009 {\it Quantum Information} (Oxford: Oxford University Press).

\bibitem{Nielsen&Chuang}
Nielsen M A and Chuang I L 2000 {\it Quantum Computation and Quantum Information} (Cambridge: Cambridge University Press.

\bibitem{state separation}
Chefles A and Barnett S M 1998 Quantum state separation, unambiguous discrimination and exact
cloning \textit{J. Phys. A: Math. Gen.} {\bf 31}, 10097--10103.

\bibitem{VAWJarxiv}
Vaccaro J A, Anselmi F, Wiseman H M and Jacobs K 2005 Complementarity between extractable mechanical work, accessible entanglement, and ability to act as a reference frame, under arbitrary superselection rules. arXiv:quant-ph/0501121v1.


\bibitem{VAWJ}
Vaccaro J A, Anselmi F, Wiseman H M and Jacobs K 2008 Tradeoff between extractable mechanical
work, accessible entanglement, and ability to act as a reference system, under arbitrary superselection rules
\textit{Phys. Rev. A} {\bf 77} 032114.

\bibitem{Arecchi}
Arecchi F T, Courtens E, Gilmore R and Thomas H 1972  Atomic Coherent States in Quantum Optics
\textit{Phys. Rev. A} {\bf 6} 2211--2237.


\bibitem{BartlettWiseman}
Bartlett S D and Wiseman H M 2003 Entanglement Constrained by Superselection Rules
\textit{Phys. Rev. Lett.} {\bf 91} 097903.

\bibitem{Rose}
Rose M E 1957 \textit{Elementary Theory of Angular Momentum} (New York: Wiley).



%\bibitem{supplementary}
%See e.g. equations (S30) and (S46) in the Supplementary Materials of Ref.~\cite{Aberg}.

%\bibitem{SWW}
%B. Schumacher, M. Westmoreland, and W.K. Wootters, Limitation on the Amount of Accessible Information in a Quantum Channel. \textit{Phys. Rev. Lett.} {\bf 76}, 3452--3455 (1996)




\end{thebibliography}
\end{document}